\documentclass[a4paper,fleqn,usenatbib]{mnras}

\usepackage{newtxtext,newtxmath}

\usepackage[T1]{fontenc}
\usepackage{ae,aecompl}

\usepackage{graphicx}	
\usepackage{amsmath}	
\usepackage{amssymb}	
\usepackage{xcolor}
\usepackage{nccmath}
\usepackage{float}
\usepackage{hyperref}
\usepackage{subcaption}

\DeclareMathOperator*{\argmax}{arg\,max}
\DeclareMathOperator*{\argmin}{arg\,min}

\title[Deblending galaxy superpositions with GANs]{Deblending galaxy superpositions with branched generative adversarial networks}

\author[D. M. Reiman and B. E. G{\"o}hre]{David M. Reiman$^{1}$\thanks{E-mail: \href{mailto:dreiman@ucsc.edu}{dreiman@ucsc.edu}}\thanks{Address: 314 Interdisciplinary Sciences Bldg., 1156 High Street, Santa Cruz, California 95064, USA}, Brett E. G{\"o}hre$^{1}$\thanks{E-mail: \href{mailto:bgohre@ucsc.edu}{bgohre@ucsc.edu}}\thanks{Address: 375 Thimann Labs, 1156 High Street, Santa Cruz, California 95064, USA}
\\
$^{1}$Department of Physics, University of California Santa Cruz, 1156 High Street, Santa Cruz, California 95064, USA}

\date{Accepted XXX. Received YYY; in original form ZZZ}

\pubyear{2018}

\begin{document}
\label{firstpage}
\pagerange{\pageref{firstpage}--\pageref{lastpage}}
\maketitle

\begin{abstract}
Near-future large galaxy surveys will encounter blended galaxy images at a fraction of up to 50\% in the densest regions of the universe. Current deblending techniques may segment the foreground galaxy while leaving missing pixel intensities in the background galaxy flux. The problem is compounded by the diffuse nature of galaxies in their outer regions, making segmentation significantly more difficult than in traditional object segmentation applications. We propose a novel branched generative adversarial network (GAN) to deblend overlapping galaxies, where the two branches produce images of the two deblended galaxies. We show that generative models are a powerful engine for deblending given their innate ability to infill missing pixel values occluded by the superposition. We maintain high peak signal-to-noise ratio and structural similarity scores with respect to ground truth images upon deblending. Our model also predicts near-instantaneously, making it a natural choice for the immense quantities of data soon to be created by large surveys such as LSST, Euclid and WFIRST.
\end{abstract}




\section{Introduction}

Large galaxy surveys necessarily encounter galactic blends due to projection effects and galactic interactions. Often, the objective of astronomical research involves measuring the properties of isolated celestial bodies. The methods used therein frequently make strong assumptions about the isolation of the object. In such cases, galactic blends can be severe enough to warrant discarding the images. Current surveys such as DES (Dark Energy Survey) \citep{DES}, KiDs (Kilo-Degree Survey) \citep{KiDS} and HSC (Hyper Suprime-Cam) \citep{HSC} as well as future surveys such as LSST (Large Synoptic Survey Telescope) \citep{BookLSST}, Euclid \citep{Euclid} and WFIRST (Wide-Field Infrared Survey Telescope) \citep{WFIRST} will generate immense quantities of imaging data within the next decade. Thus, a robust and high-throughput solution to galaxy image deblending is vital for drawing maximal value from near-future surveys. 

This issue is especially pressing given the details of LSST, for example. Beginning full operation in 2023, LSST is expected to retrieve 15 PB of image data over its scheduled 10-year operation with a limiting magnitude of $i \approx 27$. In the densest regions of the sky (100 galaxies per square arc-minute), up to half of all images are blended with center-to-center distance of 3 arc-seconds, with one-tenth of all galaxy images blended with center-to-center distance of 1 arc-second. Even considering the most typical regions of sky (37 galaxies per square arc-minute), simulation estimates suggest around 20 percent of galaxies will be superimposed at 3 arc-second separation and up to 5 percent at 1 arc-second separation. \citep{Fraction}. At 1.5 PB per year and image sizes of approximately 6 GBs, conservative estimates predict the LSST capable of producing 200,000 wide-field images. This results in 1 Billion postage stamp galaxy images per year \citep{BookLSST}. Improved handling of blended images can thus be expected to save up to 200 Million galaxy images from being discarded each year. 50 Million of these galaxy images are predicted to be blended with 1 arc-second center-to-center separation, posing an extremely difficult task for deblending algorithms. Moreover, detecting blends in the first place will pose an additional challenge for LSST (and other near-future surveys), especially for galaxies blended at 1 arc-second separation and a median seeing of ${\sim}0.67$ arc-seconds.

Pioneering work in deblending can be traced back to Jarvis \& Tyson's "Faint Object Classification and Analysis System" \citep{FOCAS} and Irwin's "Automatic Analysis of Crowded Fields" \citep{Irwin}. In Jarvis \& Tyson's \textsc{focas} algorithm, objects are initially identified and segmented by comparing the local flux density to that of the average sky density computed in a 5-by-5 pixel box filter. Blended sources are then separated by scanning the principle axis of the object centroid for a multi-modal intensity signal; if found, objects are deblended by drawing a boundary perpendicular to the principle axis at the point of minimum intensity.  Meanwhile, in Irwin's analysis, a maximum likelihood parameter estimation scheme was used to segment stars near the center of globular clusters. The technique iteratively estimates the local sky background then divides the field into images and analyzes each image for blends, estimates source position and shape, updates the background estimate and repeats. Irwin's method offered great progress in regions of high number density where the majority of images overlap even at high isophotes.

Since Jarvis \& Tyson and Irwin, many wide-field image processing algorithms have been proposed including \textsc{next} \citep{NExt} and the widely used \textsc{sextractor} \citep{SExtractor}. \textsc{next} was among the first wide-field image processing methods to utilize artificial neural networks, approaching the problem of detection as one of clustering then employing a modified version of the \textsc{focas} algorithm to deblend superimposed objects. First, a neural network compresses windows of the input image into a dense vector, performing a non-linear principal component analysis. These representations are passed through a second network which classifies the central pixel in the window as belonging to an object or the background. Neighboring object pixels are grouped together and parameters such as the photometric barycenter and principal axis are computed for each contiguous cluster. Building upon the \textsc{focas} algorithm, \textsc{next} then searches for multiple peaks in the light distribution along the principal axis as well as five other axes rotated by up to $\pi$ radians from the principal axis. When two peaks are found in the distribution, the objects are split along a line perpendicular to the axis joining the peaks. In this way, the \textsc{next} deblending algorithm is an extension of \textsc{focas} in which the assumption that the multi-peaked light distribution will occur along the principal axis is relaxed.

Modern approaches to deblending include gradient and interpolation-based deblending (\textsc{gain}) \citep{GAIN}, morpho-spectral component analysis (\textsc{muscadet}) \citep{Muscadet} and constrained matrix factorization (\textsc{scarlet}) \citep{Scarlet}. In both \textsc{muscadet} and \textsc{scarlet}, each astronomical scene is assumed to be the composition of two non-negative matrices which encode the spectral energy distribution and spatial shape information of a finite number of components which sum to represent the scene. Though the techniques share many similarities, the more recent approach of \textsc{scarlet} can be viewed as a generalization of \textsc{muscadet} which allows any number of constraints to be placed on each source. Meanwhile, \textsc{gain} acts as a secondary deblender to repair flawed images by making use of image intensity gradient information and interpolating the missing flux from background sources after the foreground object is segmented. While \textsc{gain}, \textsc{muscadet} and \textsc{scarlet} appear to be powerful deblending algorithms in their own right, here we present a method inspired by promising results in computer vision and deep learning.

Deep convolutional neural networks have proven highly effective at classifying images of galaxies \citep{Sander}. Moreover, their use as generative models for galaxy images has yielded impressive results. In particular, generative adversarial networks (GANs) with deep convolutional generators are able to model realistic galaxies by learning an approximate mapping from a latent space to the distribution over galaxy images given large datasets such as those provided by Galaxy Zoo \citep{Forging, GalaxyZoo}.

 Here, we introduce an algorithm which offers progress on both the task of deblending conspicuously blended galaxies as well as the challenge of fast, end-to-end inference of galaxy layers: a branched generative adversarial network. GANs offer an elegant solution to the problem of deblending by allowing us to combine the standard content loss function of a convolutional neural network (e.g. mean squared error) with the adversarial loss signal of a second ``discriminator'' network which pushes solutions to the galaxy image manifold and thereby ensures that our deblending predictions appear to be galaxy-like in a probabilistic sense.
 
 After training our deblending network, a forward pass of input to deblended outputs takes a trivial amount of time on a modern laptop. We focus on galaxy pairs which are known to be blended though future work may involve the identification of blends and inference of the number of galaxies involved. With this work, we intend to make usable hundreds of millions more galaxy images, while saving time and effort for astronomers to spend on other pertinent questions.

\section{Methods}

Generative adversarial networks (GANs) are a family of unsupervised learning models used to learn a map between two probability distributions \citep{GAN, GAN2}. Often, these probability distributions live in high-dimensional space. For example, images are realizations of a high-dimensional joint probability distribution over the pixel intensities. Instead of positing a parametric family of functions and using maximum likelihood methods to determine an analytic form for these complex distributions, GANs use an artificial neural network to learn the map between a known (hereafter latent) probability distribution and the target data distribution. This map is called the generator. The generator is trained by a second network which learns to discriminate between samples from the target distribution and samples from the generator's approximate distribution. This second network is called the discriminator. The generator and discriminator share a loss function in which the discriminator aims to maximize its ability to discriminate between true and generated samples while the generator aims to maximize its ability to generate samples indistinguishable from the true distribution samples and therefore "trick" the discriminator. This back-and-forth training can be viewed as a zero-sum game between neural networks where the goal is to reach Nash equilibrium \citep{GAN}. It is important to note that the generator relies upon the discriminator's signal (and hence its accuracy) to adjust its parameters and increase the quality of its generated outputs. For this reason, it is crucial that both networks train in tandem with one network not overpowering the other.

In the problem of deblending images of galaxies, we treat the joint distribution over pixel intensities of the blended images as the latent distribution. Samples from the latent distribution are denoted $I^{BL}$. The generator's purpose is to map samples from this latent distribution to corresponding samples in the deblended galaxy image distribution. Samples from the target distribution are denoted $I^{PB}$, preblended galaxy images. Our goal is to train a generator using a training set of blended images (see Data section below), $I^{BL}$, and their corresponding original images, $I^{PB}$. The discriminator provides gradient information to the generator to ensure that the generated deblended images lie on the natural image manifold. The training of the generator is also guided with supervision in the form of a pixel-wise mean squared error (see Loss section below) which further coaches the generator to closely reproduce explicit ground truth galaxy images used to make an artificial blend. After training, the generator is capable of inferring the unseen independent layers of galaxy images with high fidelity (see Results section below).

\subsection{Architecture}

\begin{figure*}
    \centering
    \includegraphics[width=\textwidth]{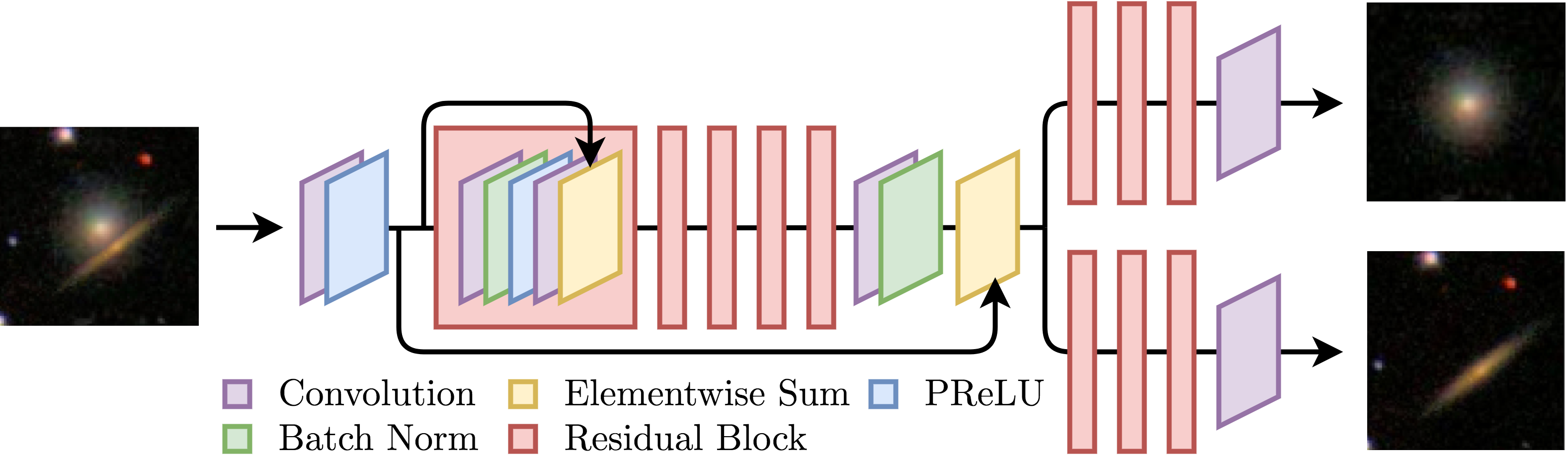}
    \caption{Branched residual network architecture for our deblender GAN. The number of residual blocks in the ``root'' and ``branches'' are integer-valued hyperparameters we denote M and N, respectively. Larger values of M and N generally produce higher quality results at the cost of training time. Here, we show (M, N) = (5, 3) as an example though the trained deblender herein used (M, N) = (10, 6). Although all residual blocks share a common structure and skip connection, here we only explicitly show the inner-workings of the first.}
    \label{fig:generator}
\end{figure*}

The architecture of our GAN is a modified version of the successful SRGAN (super resolution GAN) architecture \citep{SRGAN}. Our generator is a very deep, branched residual network \citep{ResNet} consisting of a large number of residual blocks and skip-connections (see \autoref{fig:generator}). We reference the generator as $G$ and denote the learnable parameters of the generator network as $\theta_G$. Residual networks (ResNets) have performed impressively on a variety of image detection and segmentation problems. The ResNet architecture alleviates the vanishing gradient problem by including a large number of skip-connections wherein gradient information can flow undegraded to shallow network layers, allowing for very deep network architectures. The building blocks of residual networks are residual blocks: a standard convolution with kernel size $k=[3, 3]$ followed by batch normalization \citep{BatchNorm}, activation, another convolution, batch normalization and finally an elementwise sum with the original input of the block. We refer to the first M residual blocks as the ``root'' layers whereafter the network splits into two ``branches'' with N residual blocks each. For the results presented herein, we have chosen (M, N) = (10, 6). Larger values of (M, N) are generally preferable though training times can be prohibitively expensive for large values of either. For activation functions we've chosen parametric rectified linear units (PReLU), an adaptation of leaky ReLU wherein the slope parameter  $\alpha$ becomes a learnable parameter \citep{PReLU}.

\begin{figure*}
    \centering
    \includegraphics[width=0.8\textwidth]{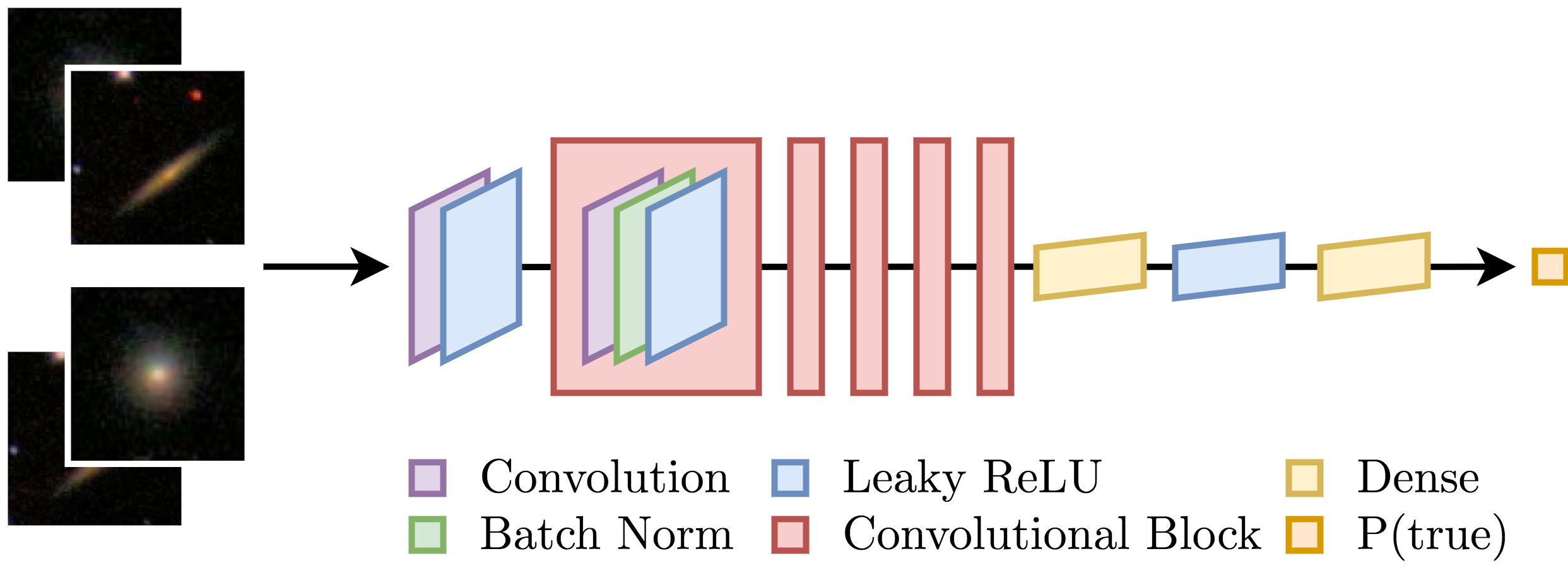}
    \caption{Deep convolutional network architecture for the deblender GAN discriminator. All convolutions use a kernel size of (3, 3) whereas strides alternate between (2, 2) and (1, 1). The number of convolutional blocks is a hyperparameter though it is constrained by the requirement that images maintain integer dimensions given that non-unity convolutional strides reduce this dimensionality by a factor of the stride assuming zero padding. Although all convolutional blocks share a common structure, here we only explicitly show the inner-workings of the first.}
    \label{fig:discriminator}
\end{figure*}

The discriminator is a deep convolutional neural network consisting of many convolutional blocks (see \autoref{fig:discriminator}). A single convolutional block consists of a convolution with kernel size $k = [3, 3]$ and unit stride followed by batch normalization and an activation layer; we employ standard leaky ReLU activations throughout with $\alpha=0.2$. We reference the discriminator as $D$ and denote the learnable parameters of the discriminator as $\theta_D$.  

For network regularization, we rely on our batch normalization layers in place of common alternatives such as dropout. In essence, batch normalization layers regularize by multiplying the hidden units by a random value (one over the batch standard deviation) and subtracting a random value (the batch mean). Since these values are different for each mini-batch, the layers must learn to be robust to added noise. 

We used TensorFlow \citep{TensorFLow} to build and run our computational graph; training was executed on a single Nvidia GTX 1080 Ti Founder's Edition video card with 11 GB of VRAM and 3584 Nvidia CUDA cores.

\subsection{Loss}

The loss function of the generator is made of three components. As in the SRGAN prescription, these losses can be broken up into two classes: (i) the adversarial loss and (ii) the content loss.

\subsubsection{Adversarial Loss}

The adversarial loss is based on the discriminator's ability to differentiate the generator's samples from true data samples. In essence, this part of the loss ensures the generator's samples lie on the galaxy image manifold (i.e. they look convincingly like galaxies). We define $p_{\text{data}}$ and $p_{z}$ as the probability distributions over the pre-blended and blended image sets, respectively, wherein $x$ is a single pre-blended image and $z$ is a single blended image.

\begin{ceqn}
\begin{equation}
l_{ADV}= \mathbb{E}_{x\sim p_{\text{data}}(x)}[\log D(x)] + \mathbb{E}_{z\sim p_z(z)}[\log(1-D(G(z)))]
\end{equation}
\end{ceqn}
\begin{ceqn}
\begin{align}
\hat{\theta}_G &= \argmin_{G} l_{ADV}(D,G)\\
\hat{\theta}_D &= \argmax_{D} l_{ADV}(D,G)
\end{align}
\end{ceqn}

In deblending use cases, this loss alone is insufficient since we aim to faithfully deblend an image into its exact components (not simply any two images that look like galaxy images). To ensure this, we employ content losses. 

\subsubsection{Content Loss}

We compute the pixel-wise mean squared error (MSE) between the generator's sample ($I^{DB}$) and the corresponding ground truth images ($I^{PB}$). This is equivalent to maximizing the effective log-likelihood of the data (in this case pixel intensities) conditional upon the network parameters given a Gaussian likelihood function. This is justified by a combination of the central limit theorem and the observation that pixel intensities arise from a very large amount of uncertain physical factors. Thus, for images of pixel width W and pixel height H, the pixel-wise mean squared error is given by the following.

\begin{ceqn}
\begin{equation}
l_{MSE} = \frac{1}{WH} \sum_{x=1}^W \sum_{y=1}^H \big[ I_{x,y}^{PB} - G_{\theta_G}(I^{BL})_{x,y} \big]^2
\end{equation}
\end{ceqn}

The second component of the content loss uses deep feature maps of the pre-trained VGG19 19-layer convolutional neural network \citep{VGG}. We chose the activations of the deepest convolutional layer (corresponding to the fourteenth overall layer of VGG19). The ground truth and generator output images are passed through the VGG19 network and their corresponding feature maps at the last convolutional layer are extracted. We compute the pixel-wise mean squared error between these feature maps and include this in the total generator loss function. For feature maps of pixel width W and pixel height H, the VGG mean squared error loss is computed as follows.

\begin{ceqn}
\begin{equation}
l_{VGG} = \frac{1}{WH} \sum_{x=1}^W \sum_{y=1}^H \big[VGG_{14}(I^{PB})_{x,y} - VGG_{14}(G_{\theta_G}(I^{BL}))_{x,y}\big]^2
\end{equation}
\end{ceqn}

In the original SRGAN implementation, the VGG feature maps were used as a content loss which resulted in perceptually satisfying images at the cost of exact similarity to the ground truth image. For our purposes, we employ the VGG loss as a ``burn-in'' loss function to break out of otherwise difficult to escape local minima. In the case of galaxy images, these local minima resulted in entirely black images for the first few hundred thousand batches. Using the VGG loss, we were able to escape the local minima in fewer than a thousand batches. To explain this improvement, note that entirely black images share much in common with images of galaxies in the dark surrounding sky, but those same flat dark images have no meaningful structural features. Because the VGG loss sets the target as deep layers of a CNN trained for image processing tasks, it effectively measures and penalizes the lack of pattern found in the predominately black images. 

Following the SRGAN prescription, we scale the adversarial loss by $10^{-3}$ to approximately match its magnitude to the mean squared error content loss. We used a discriminator-to-generator training ratio of one: for each iteration, both the discriminator's parameters and the generator's parameters were updated once. It should be noted that we have to compute the MSE and adversarial losses twice---once for each image output of the generator. The total loss is the average of the losses computed for each output image.

The composite loss function for the generator may be interpreted as a local \textit{Maximum a posteriori} estimator of the generator's parameters. Data likelihood is fixed at each training step for updating the generator networks weights, but the prior distribution is non-stationary. The discriminator is the stand-in prior at each local MAP step, and becomes more representative of our true prior while it is jointly trained. It is extremely challenging to hand engineer the features that describe what makes a galaxy appear real. We need to competitively train to approximate the ideal prior. The discriminator learns to better discern which weight configuration in the generator yields a mapping that creates realistic images. From a Bayesian perspective, the adversarial term can therefore be interpreted as a moving log-prior assigning more probabilistic weight to generators that can fool the increasingly trained discriminator.

\subsection{Data}

As noted in \cite{Blends}, large near-future surveys such as LSST will encounter three classes of blends: \textit{ambiguous} blends, \textit{conspicuous} blends and \textit{innocuous} blends. Here, we focus primarily on the conspicuous (confirmed blends) and innocuous (near blends) classes. Future work may include classifying ambiguous blends.

We used 141,553 images available from Galaxy Zoo \citep{GalaxyZoo} via the Kaggle Galaxy Zoo classification challenge. These are galaxy images from the sixth data release \citep{DR6} of the Sloan Digital Sky Survey (SDSS). SDSS is a survey covering nearly 26 percent of the sky and capturing photometric data in five filters, though the images used here were composite images of the $g$, $r$ and $i$ bands. Each image is of standard (i.e. non-extended) JPEG format and therefore has a bit depth of 8 bits per color channel. According to \cite{GalaxyZoo}, the images were cropped such that the resolution was $0.024R_p$ arc-second per pixel with $R_p$ the Petrosian radius of the system.

Each image is originally of dimension (H, W, C) = (424, 424, 3) where H, W and C are the height, width and channel dimensions, respectively. We first crop the center of these images to dimension (H, W, C) = (240, 240, 3) and then downsample them using a bicubic sharpener to (H, W, C) = (80, 80, 3). Leaving one image centered, we then perturb the second image by flipping it both horizontally or vertically according to a Bernoulli distribution with $p=0.5$, rotating uniformly on the interval $\theta \in [0, 2\pi]$, displacing both horizontally or vertically uniformly on the interval $dx, dy \in [10, 50]$ pixels and finally scaling log-uniformly on the interval $r \in [1/e, \sqrt{e}]$ where $r$ is the scale ratio. These images serve as our pre-blended ground truth images ($I^{PB}$); we scale their pixel values to the interval $I \in [-1, 1]$. 

The images are then blended by selecting the pixelwise max between the two images. This blending prescription was inspired by the catalog of true blended Galaxy Zoo images presented in \cite{BlendCatalog}. We find that pixelwise max blended images closely match selections from the catalog of true blends though we discuss improvements on this schema later on (see Discussion section below). We scale the blended image pixel values to the interval $I \in [0, 1]$ as recommended in the SRGAN paper. These blended images are realizations of the latent distribution of images $(I^{BL})$. 

Note that it is a necessary requirement for our method that the image is centered upon one galaxy; this spatial information informs the independent behavior of each branch. During training, the network learns that one branch corresponds to the deblended central galaxy, while the other branch corresponds to the deblended off-center galaxy. Without such a feature (which should always be possible in true blends with a simple image crop), the network may be confounded by which branch to assign to which galaxy, and for example create the same deblended galaxy image in each branch. By using a principled centering scheme, this problem is largely diminished. This scheme should be easily extended to more than two layers by deblending the center galaxy from all other galaxies, then sending the deblended output centered on a remaining galaxy back into the generator.

\subsection{Training}

During training, the generator network is presented with a batch of blended inputs ($I^{BL}$); we use a batch size of 16 samples. The generator predicts the deblended components ($I^{DB}$) of each blended input. The discriminator is given both a batch of generated deblended images and their corresponding preblended ground truth images. The discriminator updates its parameters to maximize its discriminative ability between the two distributions. Mathematically speaking, minimization of the adversarial loss is equivalent to minimization of the Jenson-Shannon divergence between the generated distribution $(I^{DB})$ and the data distribution $(I^{PB})$ whereas minimization of the mean squared error term equates to maximization of an effective Gaussian likelihood over pixel intensities. In reality, the generator is trained to minimize the sum of these components by averaging the gradient signals from each. This is done via stochastic gradient descent on the network's high-dimensional error manifold defined by its compound loss function with respect to its learnable network parameters. 

Both the generator and discriminator loss functions are optimized in high-dimensional neural network weight space via the Adam optimizer (adaptive moment estimation)  \citep{Adam}, a method for stochastic optimization which combines classical momentum with RMSProp. With momentum, Adam uses past gradients to influence current gradient-based steps, accelerating learning in valleys of the error manifold. Following the running-average style of RMSProp, it emphasizes the most recent gradients while decaying influence from much earlier ones, hence adapting better to the present loss surface curvature.

The steps down the error manifold are further scaled by a hyperparameter referred to as the learning rate (although the true rate is also naturally scaled by adaptive momentum). We choose an initial learning rate of $10^{-4}$. After 100,000 update iterations (1.6 million individual images or five epochs), we decrease the learning rate by an order of magnitude to $10^{-5}$. We then train for another 100,000 update iterations after which the network parameters are saved and can be used for inference on unseen images.

\section{Results}


To evaluate our model, we infer upon blended images withheld from the training dataset. For these blends, we have access to the to the true preblended images and therefore are able to quote relevant metrics for image comparison (see below). It should be noted that the galaxies which make up the testing set blended images have also been withheld in the creation of the training set. That is, the deblender GAN has never seen this particular blend nor the underlying individual galaxies during training. A selection of results are presented in \autoref{fig:successes} and \autoref{fig:failures} with a more extensive catalog of predictions in a GitHub repository available at \href{https://github.com/davidreiman/deblender-gan-images}{https://github.com/davidreiman/deblender-gan-images}.

We select two image comparison metrics for determining the quality of the deblended images in relation to the ground truth images: the peak signal-to-noise ratio (PSNR) and the structural similarity index (SSIM).

The peak signal-to-noise ratio is often used as an evaluation metric for compression algorithms. It is a logarithmic measure of the similarity between two images, in our case the ground truth preblended image ($I^{PB}$) and the deblended re-creation ($I^{DB}$). The equation for PSNR is written in terms of the maximum pixel intensity (MAX) of the truth image and the mean squared error (MSE) between the test and ground truth images.

\begin{ceqn}
\begin{equation}
    \text{PSNR} = 20\log_{10}(\text{MAX}) - 10\log_{10} (\text{MSE})
\end{equation}
\end{ceqn}

The structural similarity index \citep{SSIM} is a method for evaluating the perceptual quality of an image in relation to an unaltered, true image. It was invented as an alternative to PSNR and MSE methods which measure error. Instead, SSIM measures the structural information in an image where structural information is understood in terms of pixel-wise correlations. SSIM scores are measured via sliding windows of user-defined width in which the means ($\mu_x$, $\mu_y$), variances ($\sigma_x^2$, $\sigma_y^2$)  and covariance ($\sigma_{xy}$) of pixel intensities are computed in the same windowed region of each image being compared, the window in one image denoted $x$ and the other denoted $y$. Included are the constants $c_1$ and $c_2$ which stabilize the division to avoid undefined scores---here we use $c_1=0.01$ and $c_2=0.03$. A single SSIM score between two images is the mean of all windowed SSIM values across the extent of the images.

\begin{ceqn}
\begin{equation}
    \text{SSIM} = \frac{(2\mu_x\mu_y + c_1)(2\sigma_{xy} + c_2)}{(\mu_x^2 + \mu_y^2 + c_1)(\sigma_x^2 + \sigma_y^2 + c_2)}
\end{equation}
\end{ceqn}

\begin{figure}
    \centering
    \includegraphics[width=0.48\textwidth]{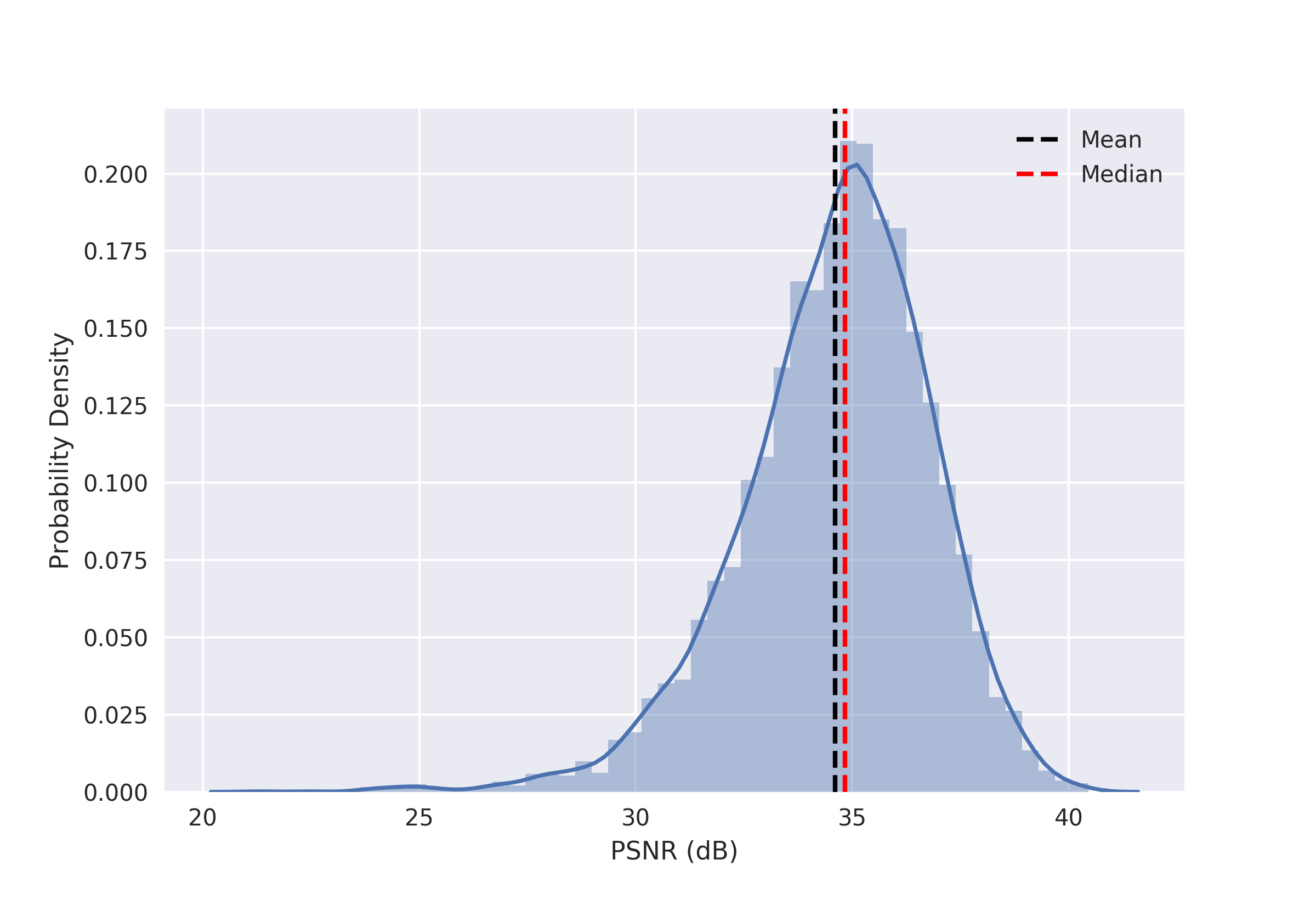}
    \caption{Distribution of PSNR scores on the testing set. Deblended images achieve mean PSNR scores of $34.61$ dB with a standard deviation of approximately $\sigma = 2.2$.}
    \label{psnr-pdf}
\end{figure}

\begin{figure}
    \centering
    \includegraphics[width=0.48\textwidth]{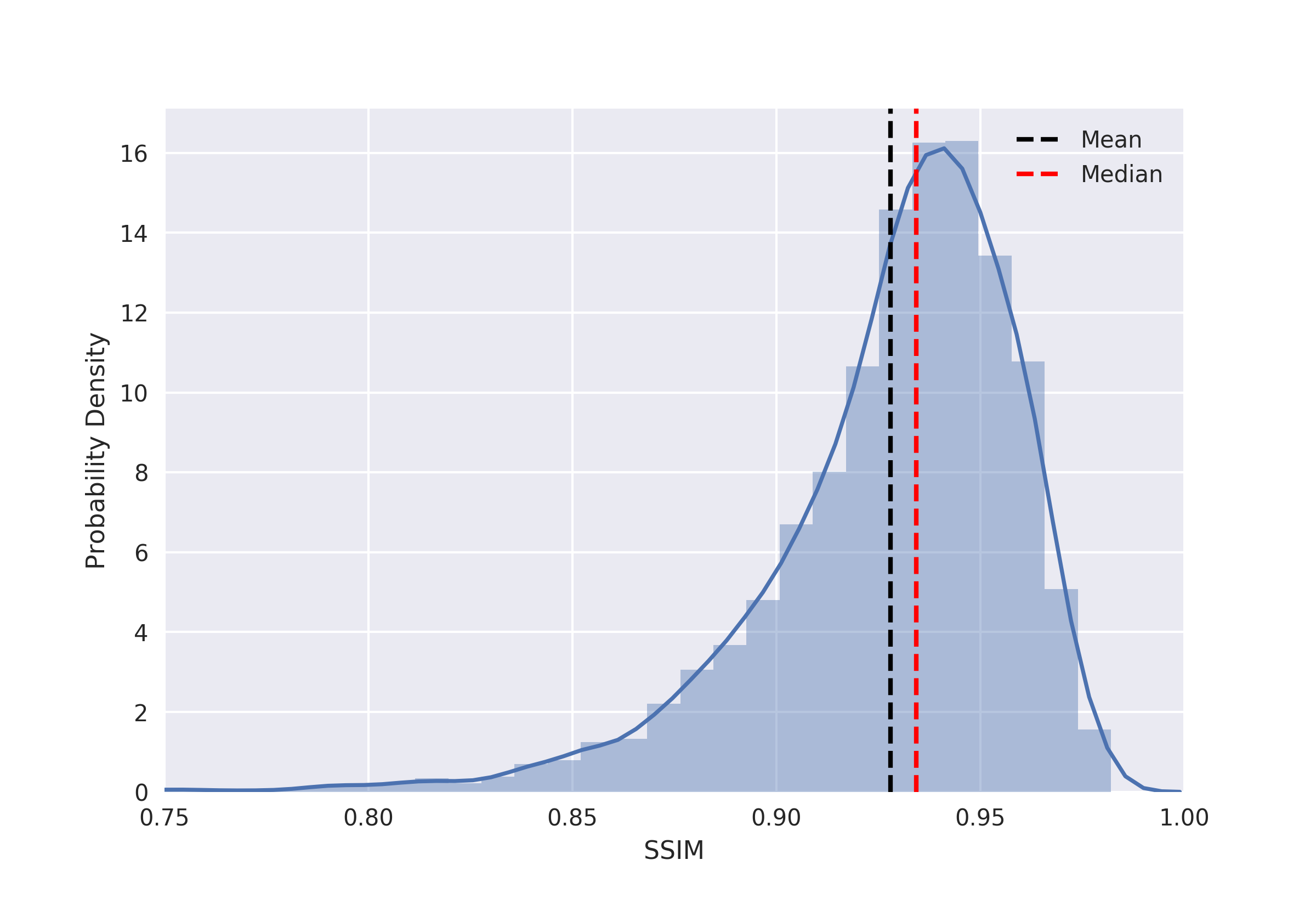}
    \caption{Distribution of SSIM scores on the testing set. Deblended images achieve a maximum structural similarity of $0.982$ with a standard deviation of approximately $\sigma = 0.032$.}
    \label{ssim-pdf}
\end{figure}

We tested on 3,200 blends withheld from the training set. This corresponds to 6,400 individual galaxy images. We computed 6,400 PSNR and SSIM scores and quote their mean, median, minimum, maximum and variance in \autoref{metric-table}. Their full distributions are displayed in \autoref{psnr-pdf} and \autoref{ssim-pdf}.

\begin{figure*}
    \includegraphics[width=0.75\textwidth]{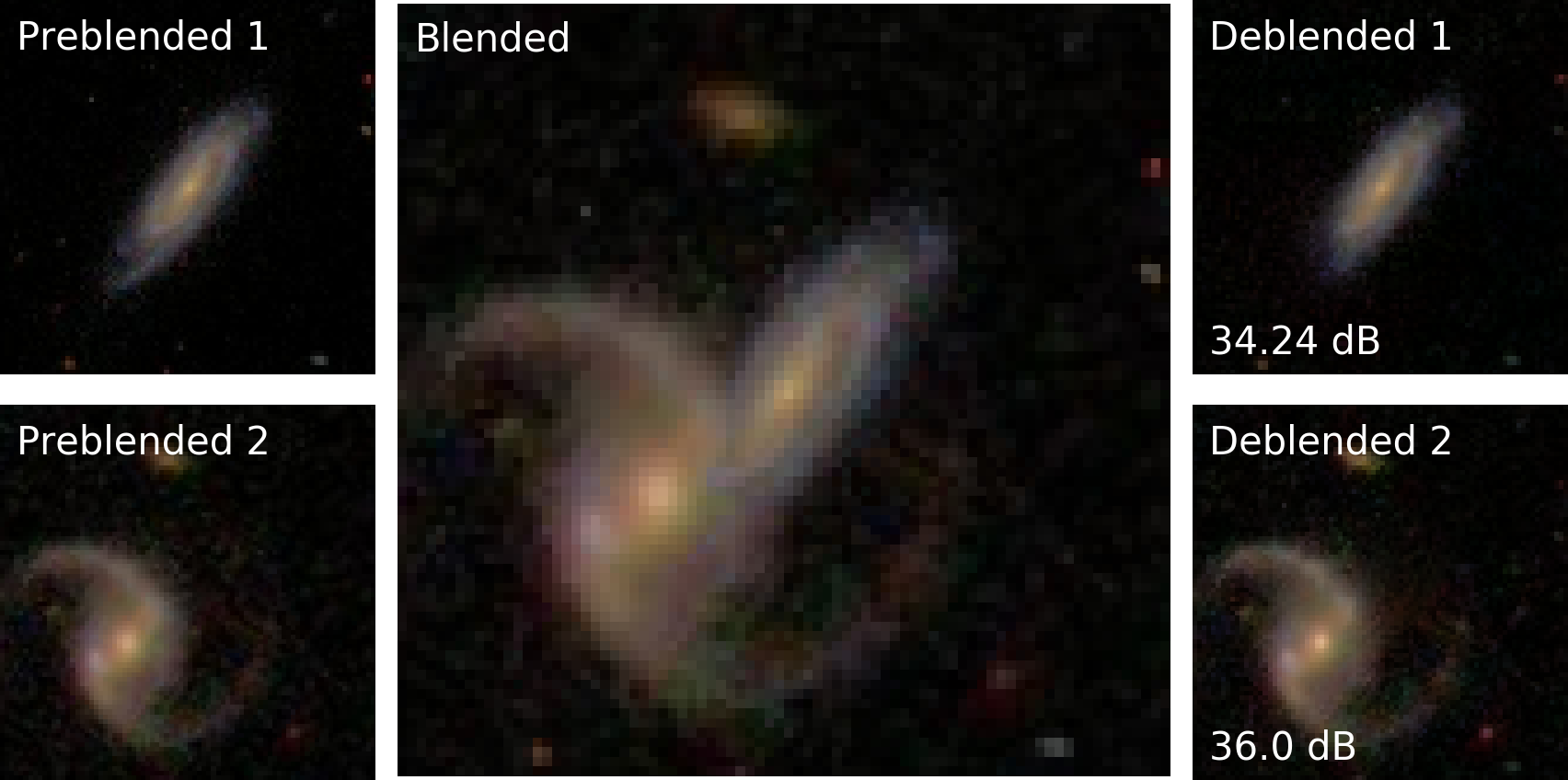}
    \includegraphics[width=0.75\textwidth]{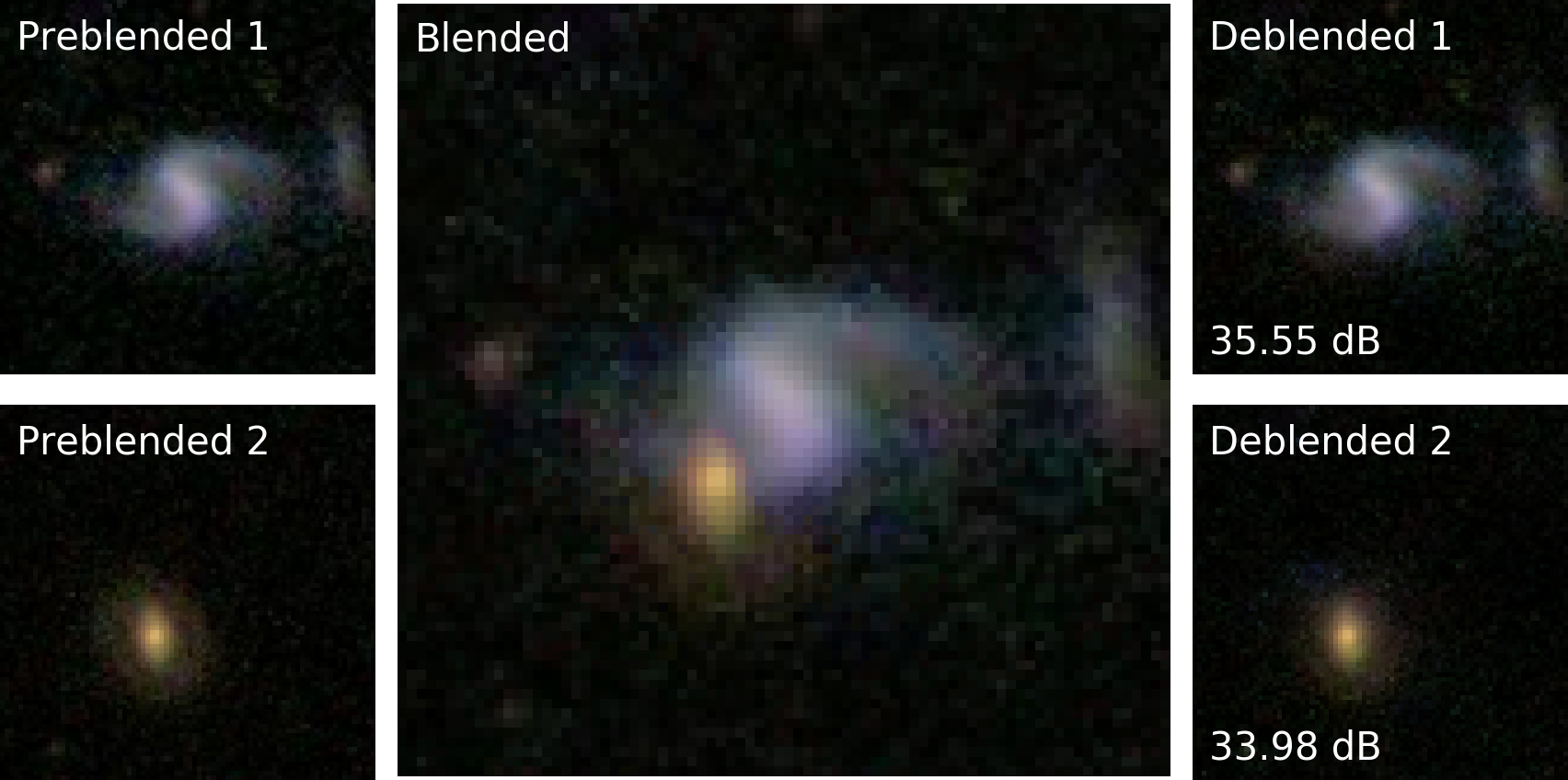}
    \includegraphics[width=0.75\textwidth]{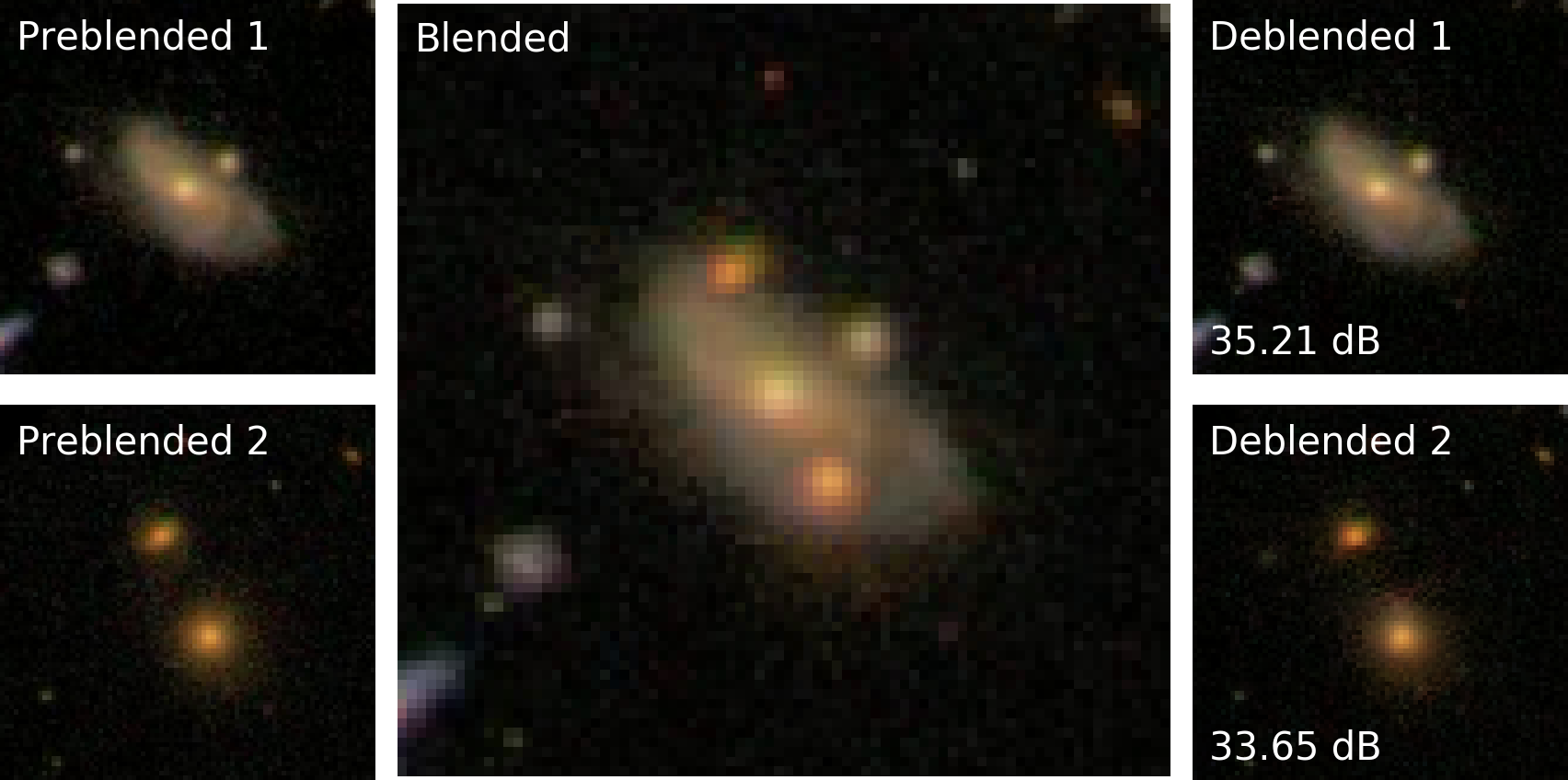}
    \caption{}
\end{figure*}
\clearpage
\begin{figure*}
    \ContinuedFloat
    \includegraphics[width=0.75\textwidth]{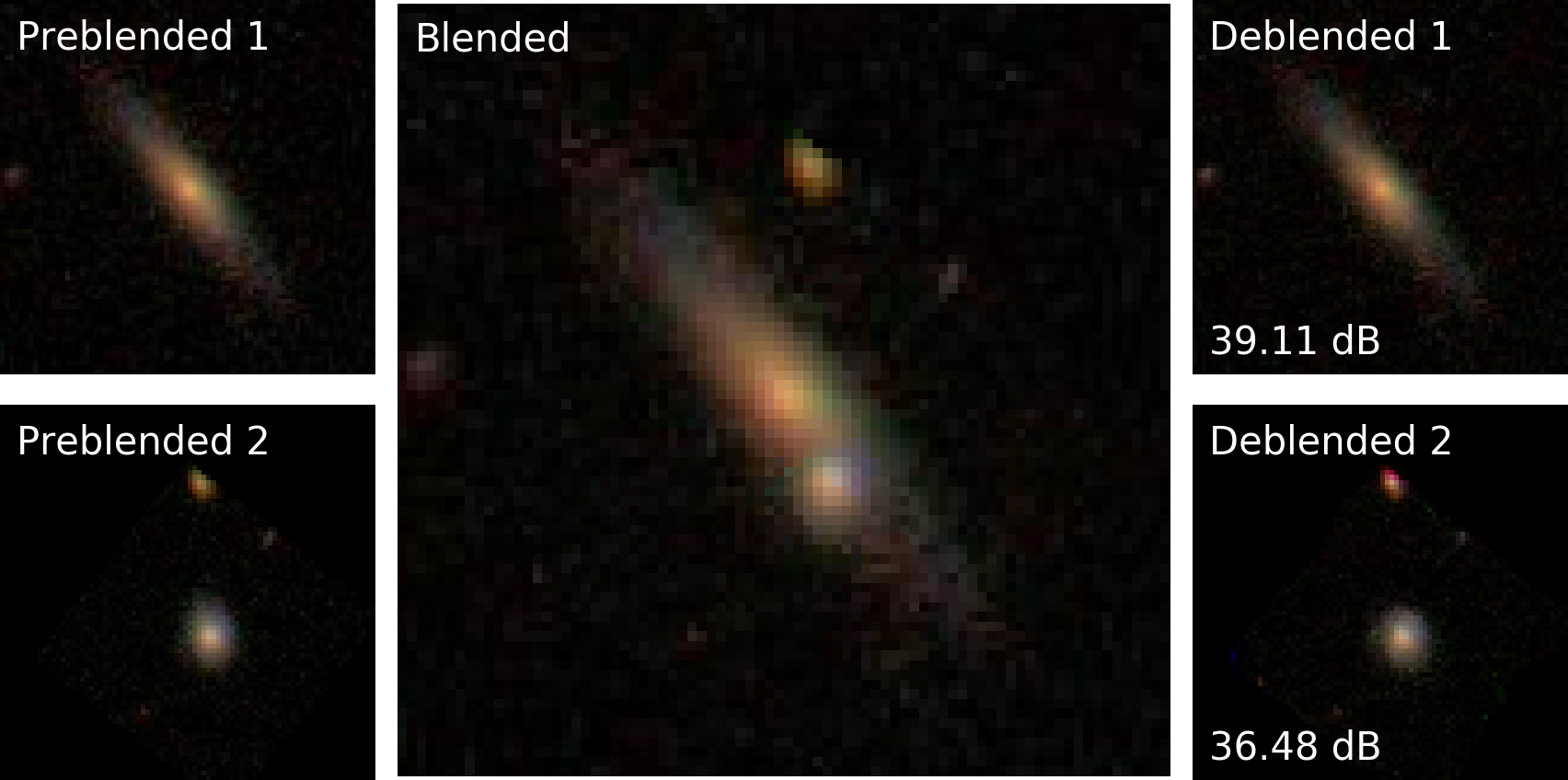}
    \includegraphics[width=0.75\textwidth]{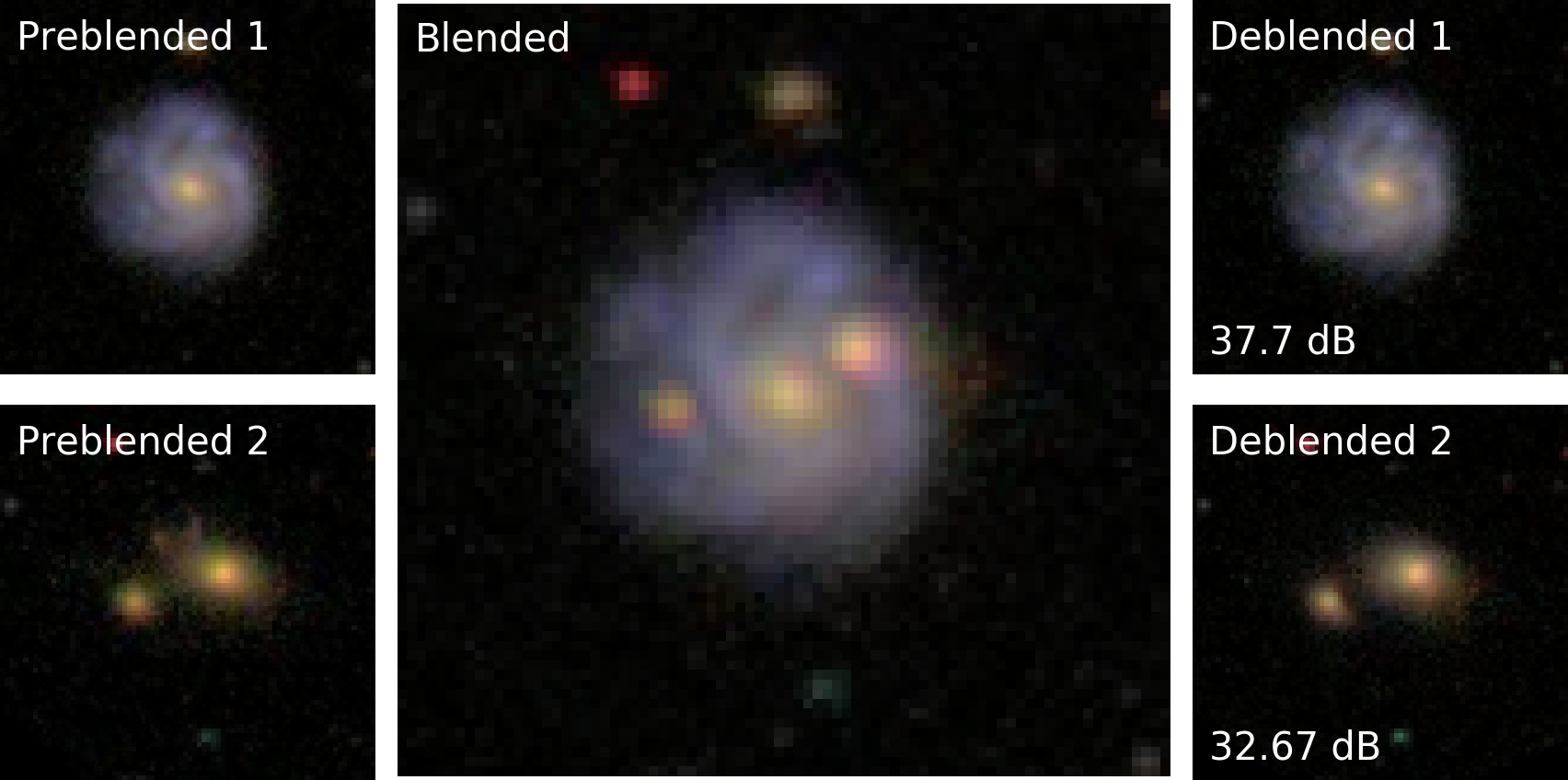}
    \includegraphics[width=0.75\textwidth]{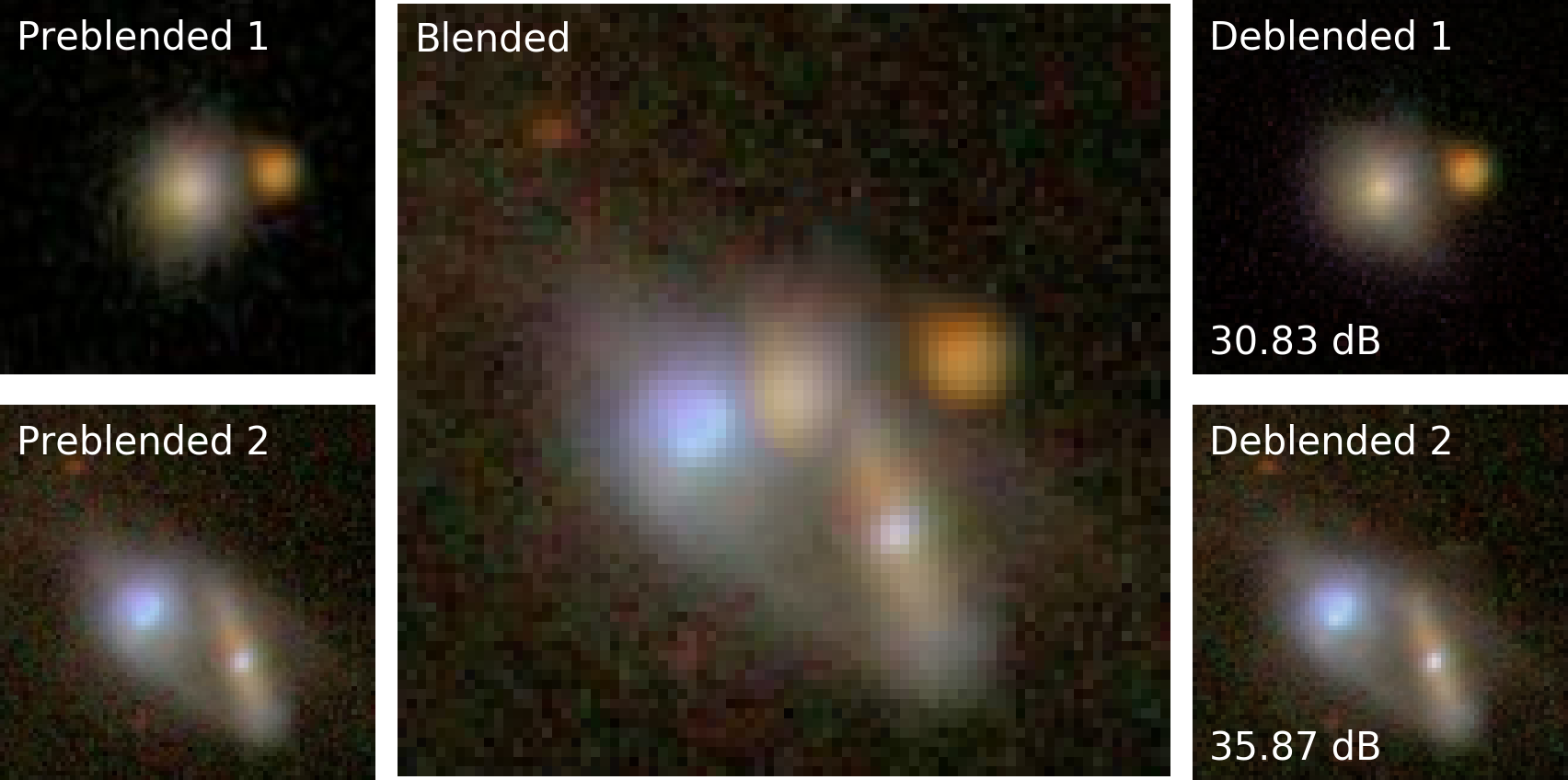}
    \caption{}
\end{figure*}
\clearpage
\begin{figure*}
    \ContinuedFloat
    \includegraphics[width=0.75\textwidth]{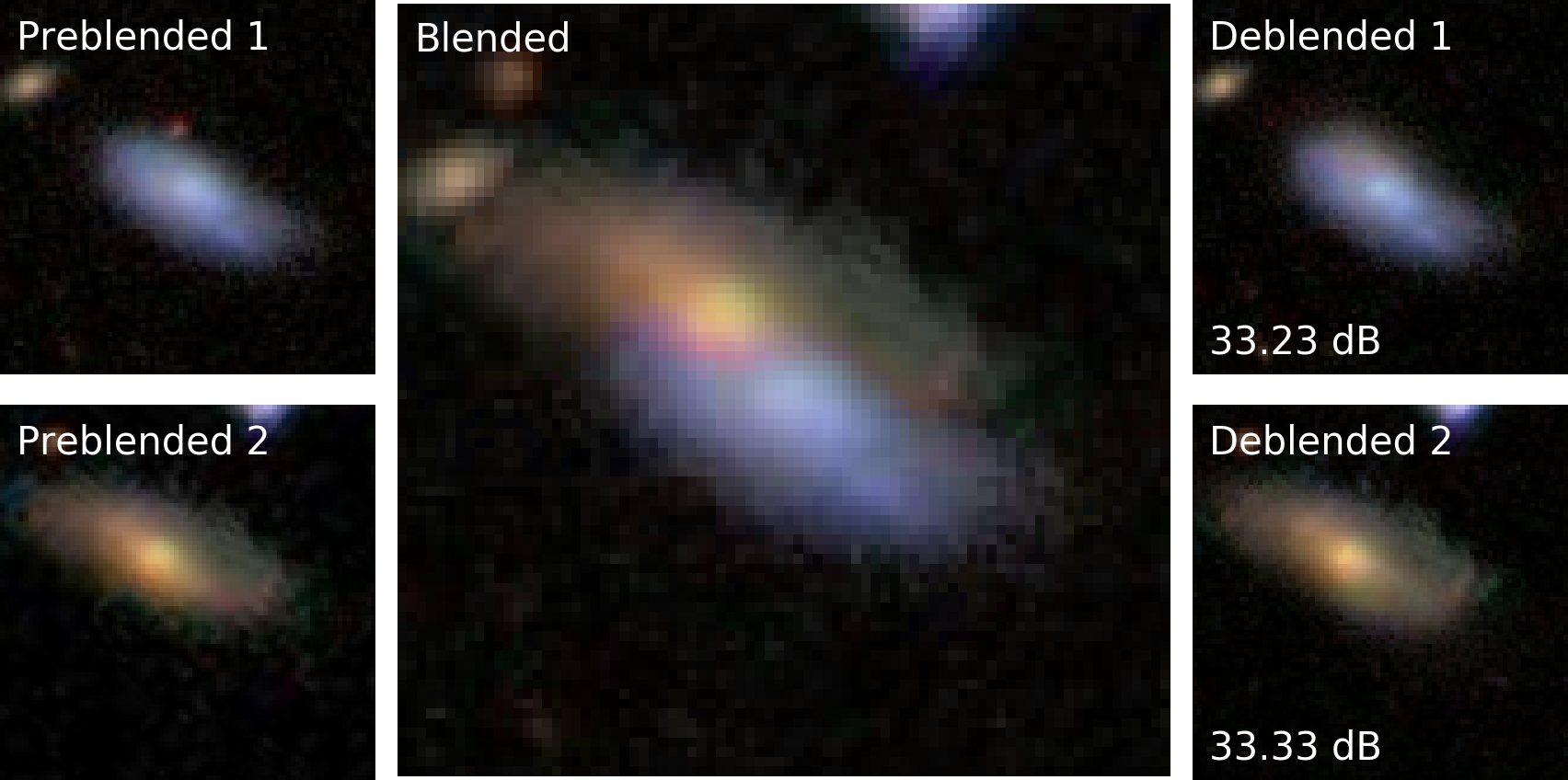}
    \includegraphics[width=0.75\textwidth]{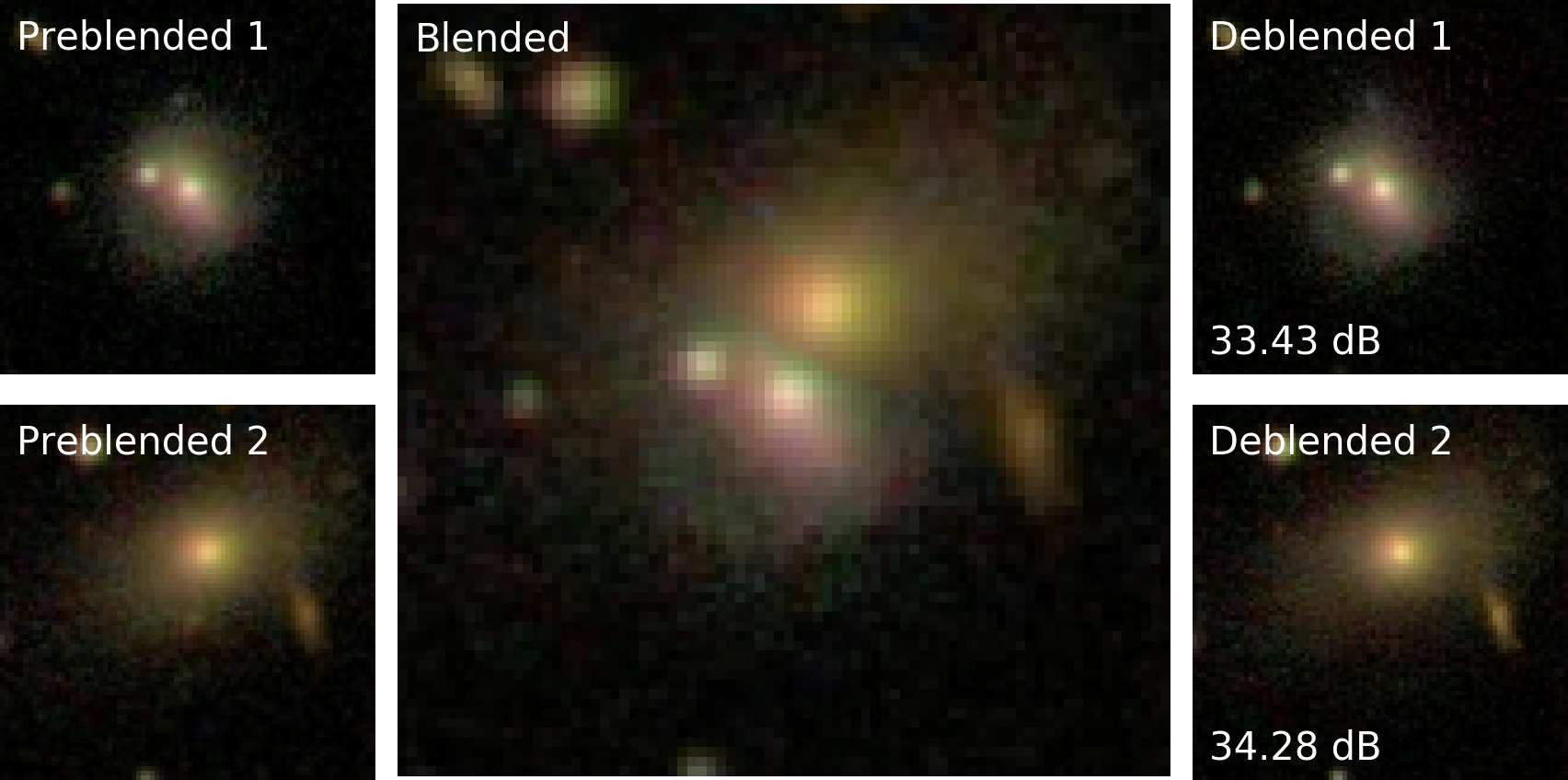}
    \includegraphics[width=0.75\textwidth]{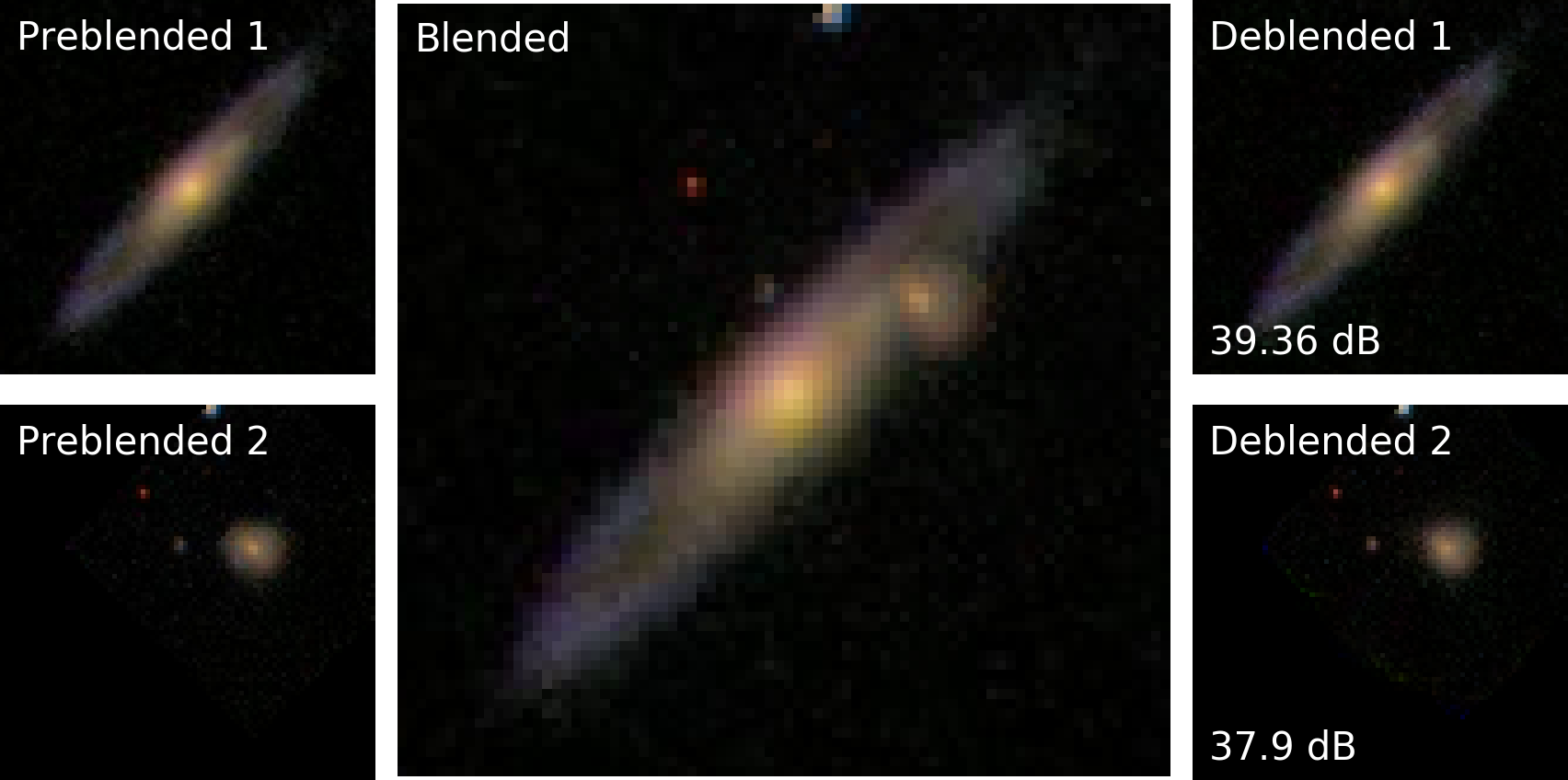}
    \caption{A selection of successful deblender GAN predictions. On the left side of each panel are two preblended Galaxy Zoo images ($I^{PB}$) which were superimposed to create the central blended input image. The trained generator's deblended predictions ($I^{DB}$) are on the right of each panel. Superimposed upon the deblended predictions are the associated PSNR scores between the deblended image and its related preblended image. A variety of galaxy morphologies are represented here---in each case, our deblender GAN successfully segments and deblends the foreground galaxy while imputing the most likely flux for the occluded pixels in the background galaxy.}
    \label{fig:successes}
\end{figure*}
\clearpage

The deblender GAN performs impressively with respect to both metrics. For reference, values of PSNR above $30$ dB are considered acceptable for lossy image compression. Considering that these images were completely reconstructed out of a blended image (rather than compressed) the scores set a high benchmark for deblending. It should be noted, however, that comparing these PSNR and SSIM scores directly to those of natural images is likely in error. Many galaxy images are dominated by a black background which can bias both the mean squared error and pixelwise correlations upward and thereby inflate the resulting PSNR and SSIM scores. Still, we quote our scores as a benchmark for comparison with future iterations of our own model and collation with alternative deblending algorithms working with a similar dataset and image format.

\begin{table}
\begin{tabular}{rccccc}
\hline
\multicolumn{1}{l}{Imaging Metrics} & Mean  & Median & Max   & Min   & \multicolumn{1}{l|}{Var} \\ \hline
PSNR (dB)                           & 34.61 & 34.83  & 40.45 & 21.34 & 4.88                     \\
SSIM                                & 0.928 & 0.934  & 0.982 & 0.576 & 0.001                    \\ \hline
\end{tabular}
\caption{Summary statistics of the PSNR and SSIM score distributions. Maximum PSNR and SSIM scores are comparable with high-quality compression algorithms ($\text{PSNR} \in [30, 50] \text{ dB}$).}
\label{metric-table}
\end{table}

\section{Discussion}

We've shown promising initial results for the application of a novel generative adversarial network architecture to the problem of galaxy deblending. Deep generative models are natural solutions to deblending for their ability to infill missing pixel values behind blends using information learned about the distribution of natural images provided during training. The discriminative loss ensures that all deblended images lie on the natural image manifold of galaxies. In addition, the branched structure allows our model to identify separate galaxies and segment them accordingly without human intervention or labeling; our approach only requires that one of the galaxies lie at the center of the image. Our branched GAN naturally handles the large incoming quantities of data from surveys like LSST and DES, deblending images near-instantaneously.

We've used the peak signal to noise ratio (PSNR) and structural similarity index (SSIM) as a metrics of quality in the segmentation. Though we've found no other applications of deep learning to the problem of deblending, we have quoted our PSNR and SSIM scores here as a benchmark for future comparisons with our own branched GAN revisions and with the alternative deblending algorithms.

We encountered a variety of issues while training our branched deblender GAN. Most notable of these is the tendency to fall into local minima near the start of training. Since galaxy images are largely black space the GAN learns that it can initially trick the discriminator by making all black images. We broke out of the minima by including a ``burn-in'' period of a few thousand batches using only a mean squared error loss between the deep activations of the pre-trained VGG19 network. Another solution to this issue may be to pre-train the discriminator though there exists the possibility of mode collapse wherein the generator and discriminator performance are highly unequal and learning ceases.

There is great room for improvement in our branched GAN. Our blending schema was chosen to match the curated catalog of true blends in the Galaxy Zoo catalog presented in \cite{BlendCatalog} though it certainly could be improved upon. In truth, galaxy blends consist of a combination of pixelwise sum (generally in more diffuse regions) and pixelwise max/min (generally where dense regions of the foreground galaxy obstructs the background galaxy flux). Pure pixelwise max blending tends to create unrealistic blended images where low flux foreground galaxies fall upon bright regions of the background galaxy---this generally leads to blends with artificially incomplete foreground galaxies which have been ``cutoff'' by the high flux pixels of the background images. Images of this type bias our estimates of both PSNR and SSIM scores. Indeed, pixelwise max blending is in some sense the worst case for deblending wherein precisely zero information about the background galaxy is encoded in the pixel intensities in regions where the foreground galaxy eclipses it. On the other hand, a variety of generated blends exhibit artificially hard lines between overlapping galaxies which likely makes the galaxies easier to deblend. Moreover, the presence of unmasked background galaxies in the Galaxy Zoo images give rise to artificial penalties in the mean squared error loss function when they are assigned to the incorrect deblended galaxy image. We also note a slight misalignment of the RGB channels of the Galaxy Zoo images which gives rise to unnatural color gradients---a feature that our network learned to reproduce. We propose to address issues in our blending schema and issues in data creation in general in future work (see below).

\begin{figure*}
    \includegraphics[width=0.75\textwidth]{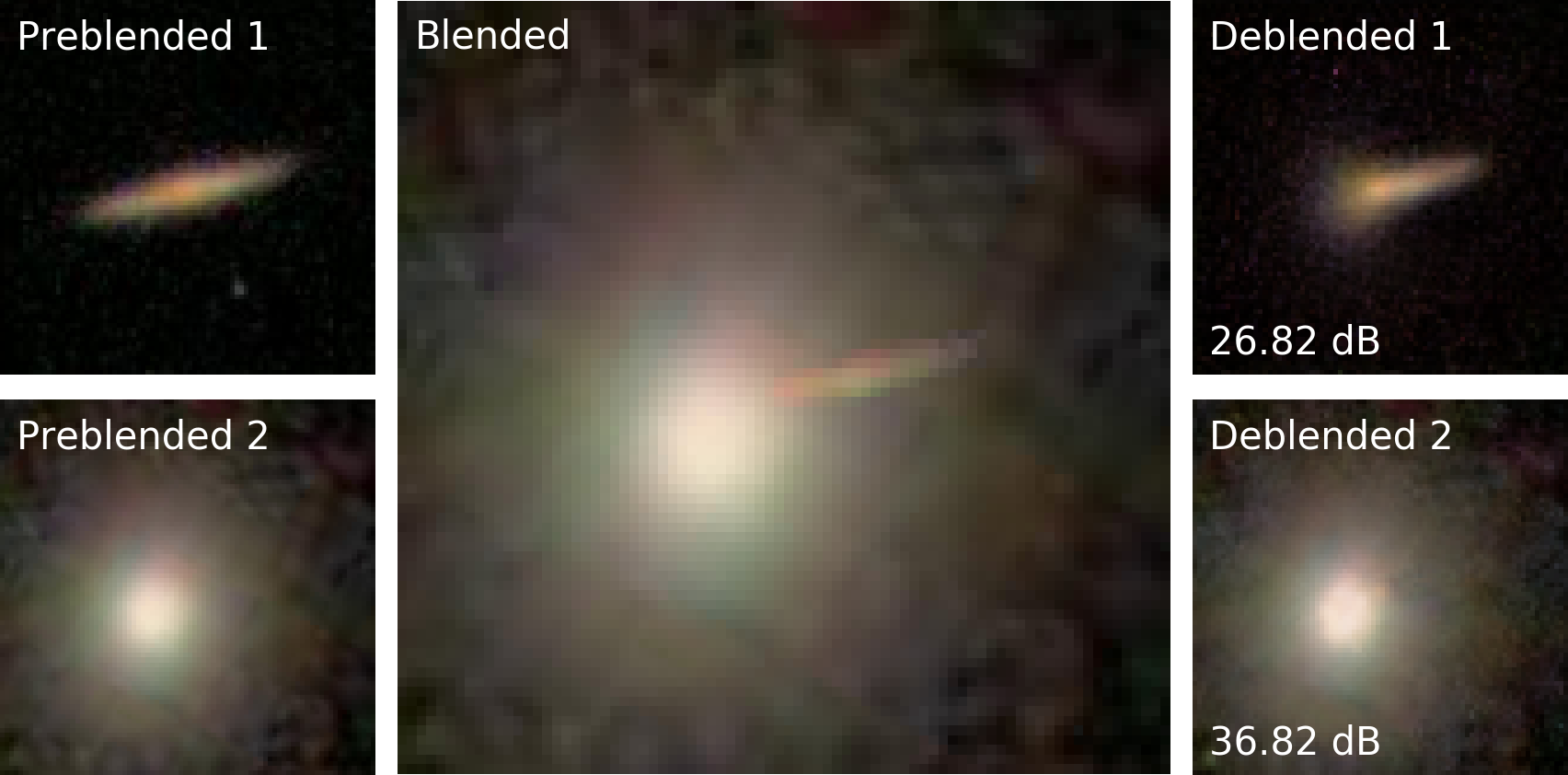}
    \includegraphics[width=0.75\textwidth]{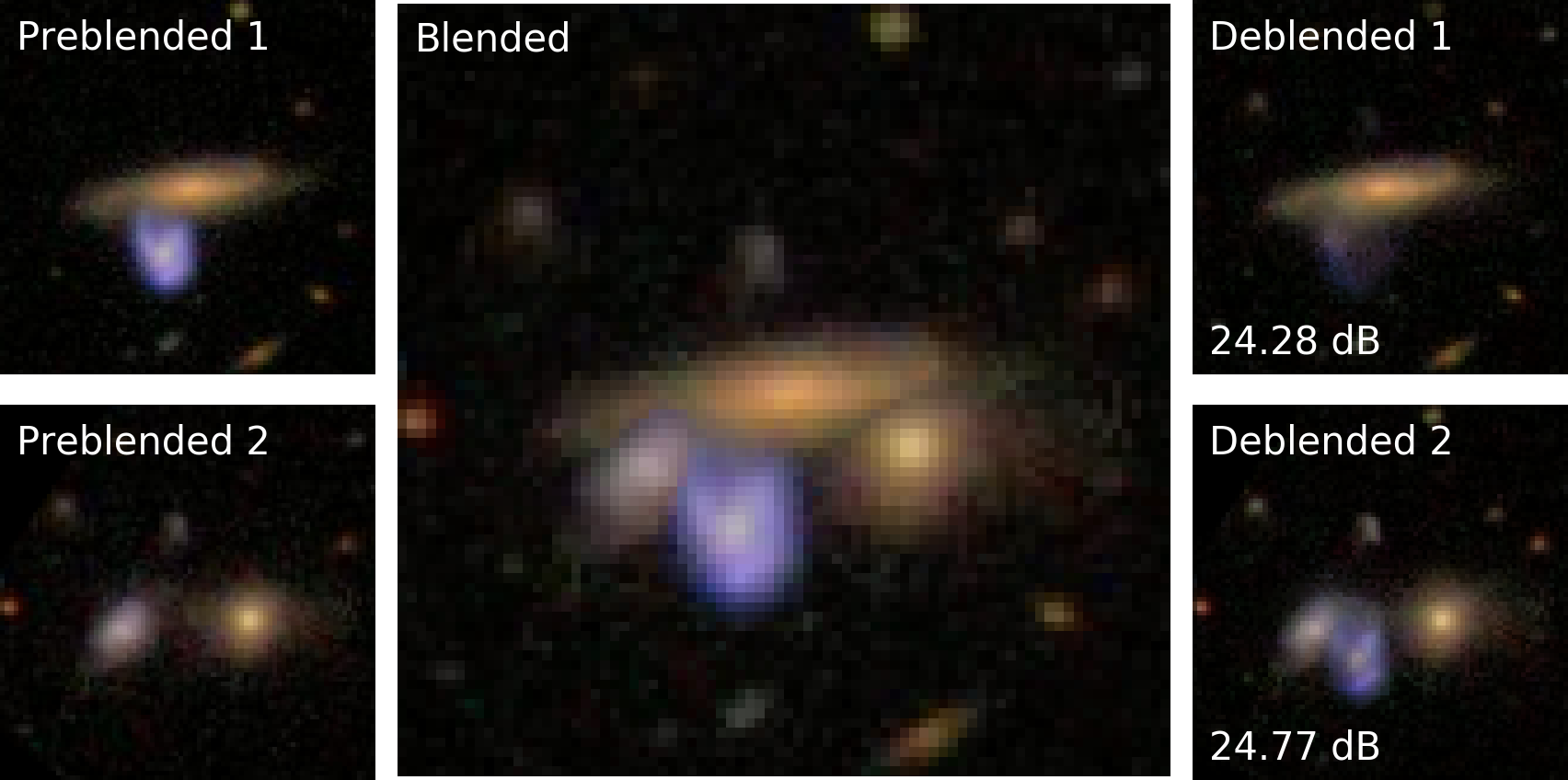}
    \includegraphics[width=0.75\textwidth]{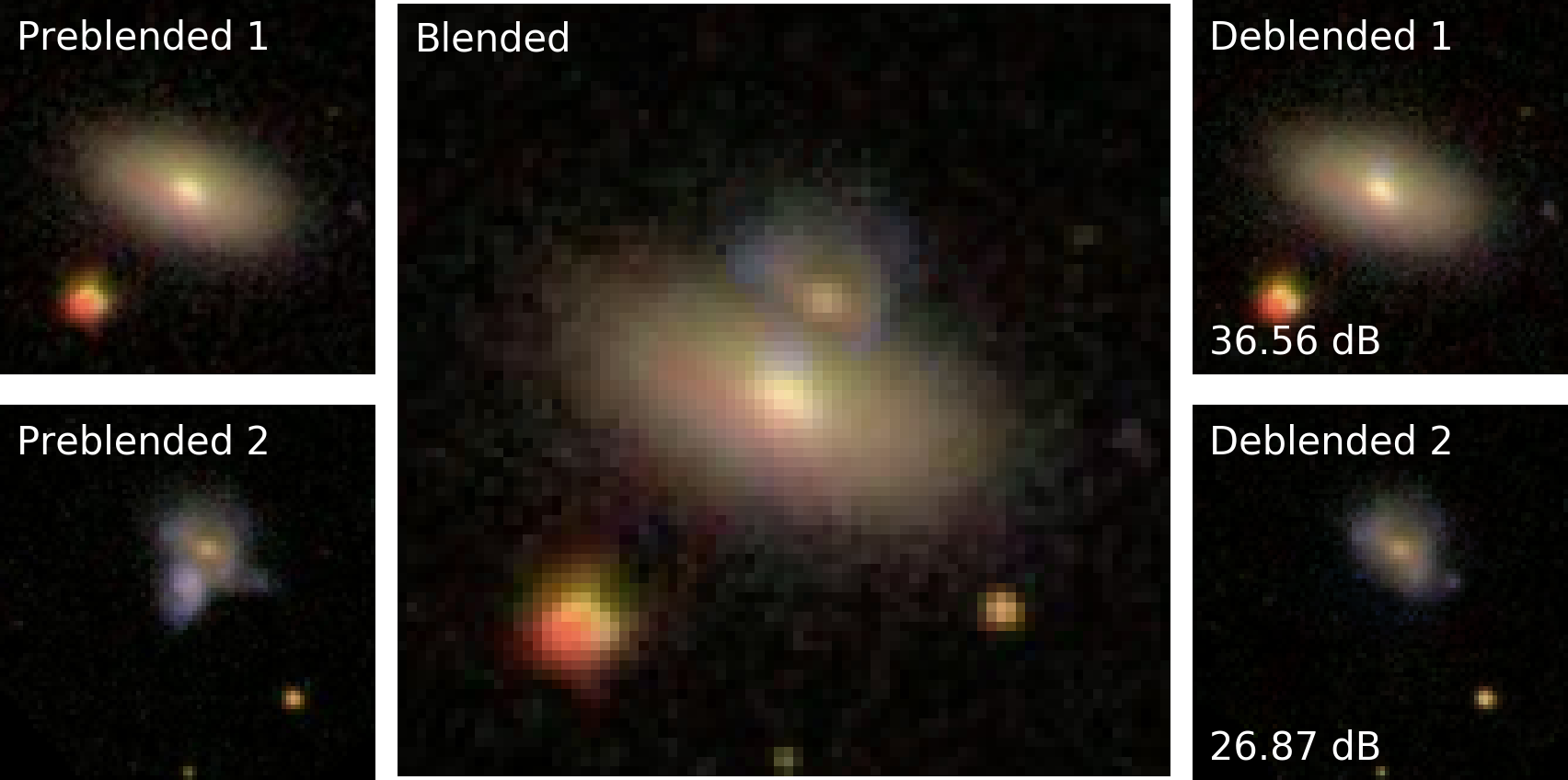}
    \caption{}
\end{figure*}
\clearpage
\begin{figure*}
    \ContinuedFloat
    \includegraphics[width=0.75\textwidth]{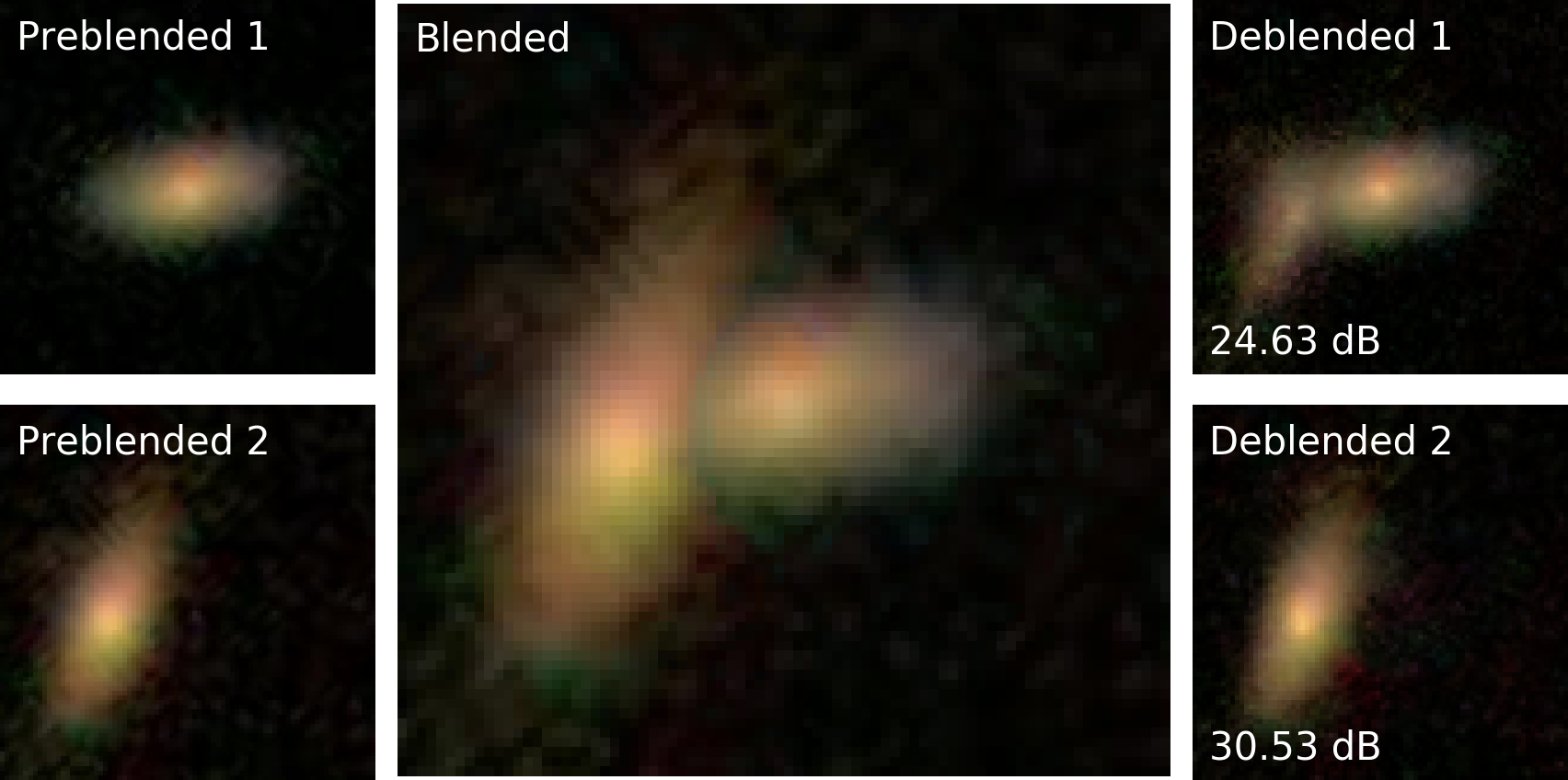}
    \includegraphics[width=0.75\textwidth]{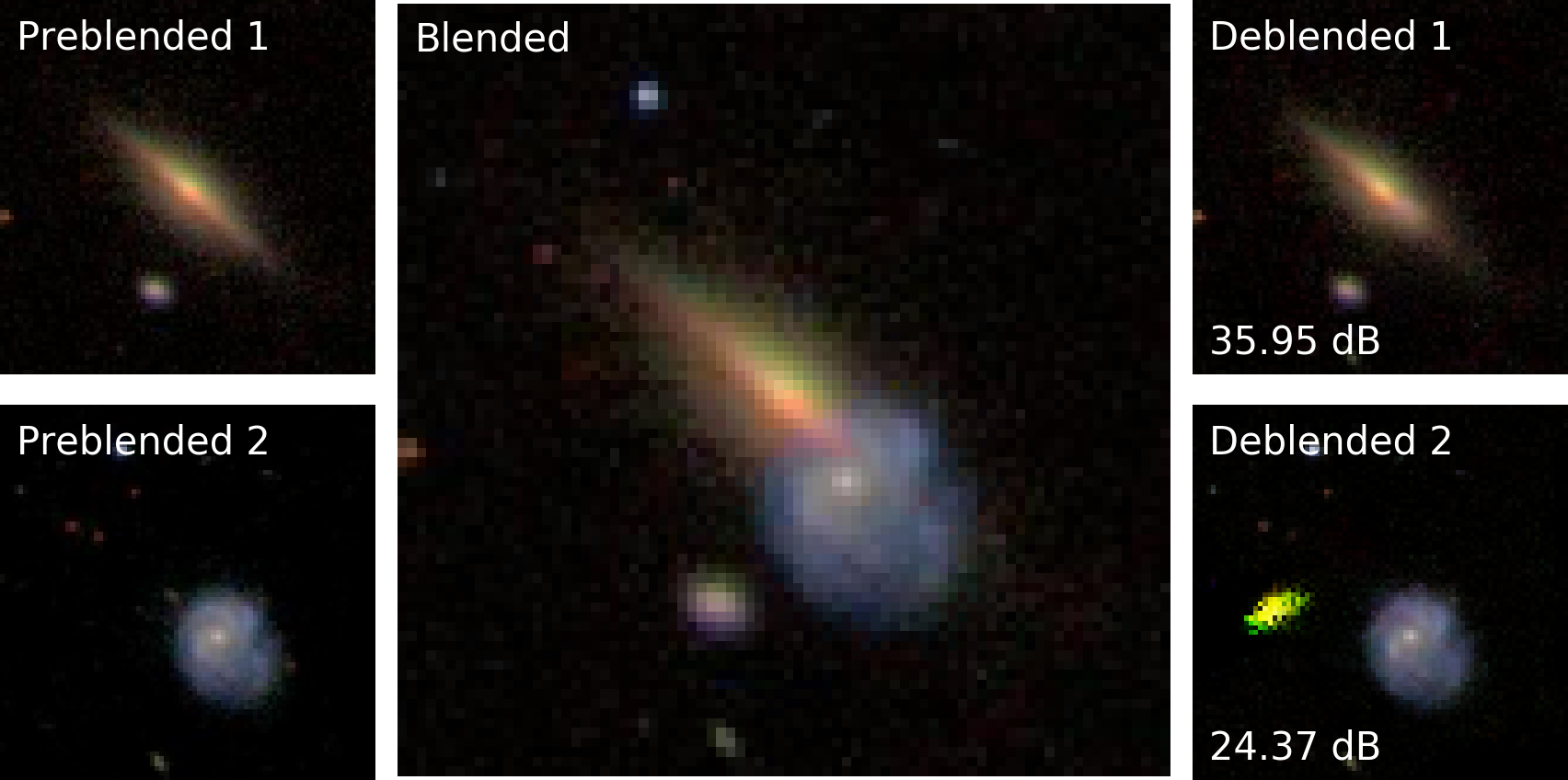}
    \includegraphics[width=0.75\textwidth]{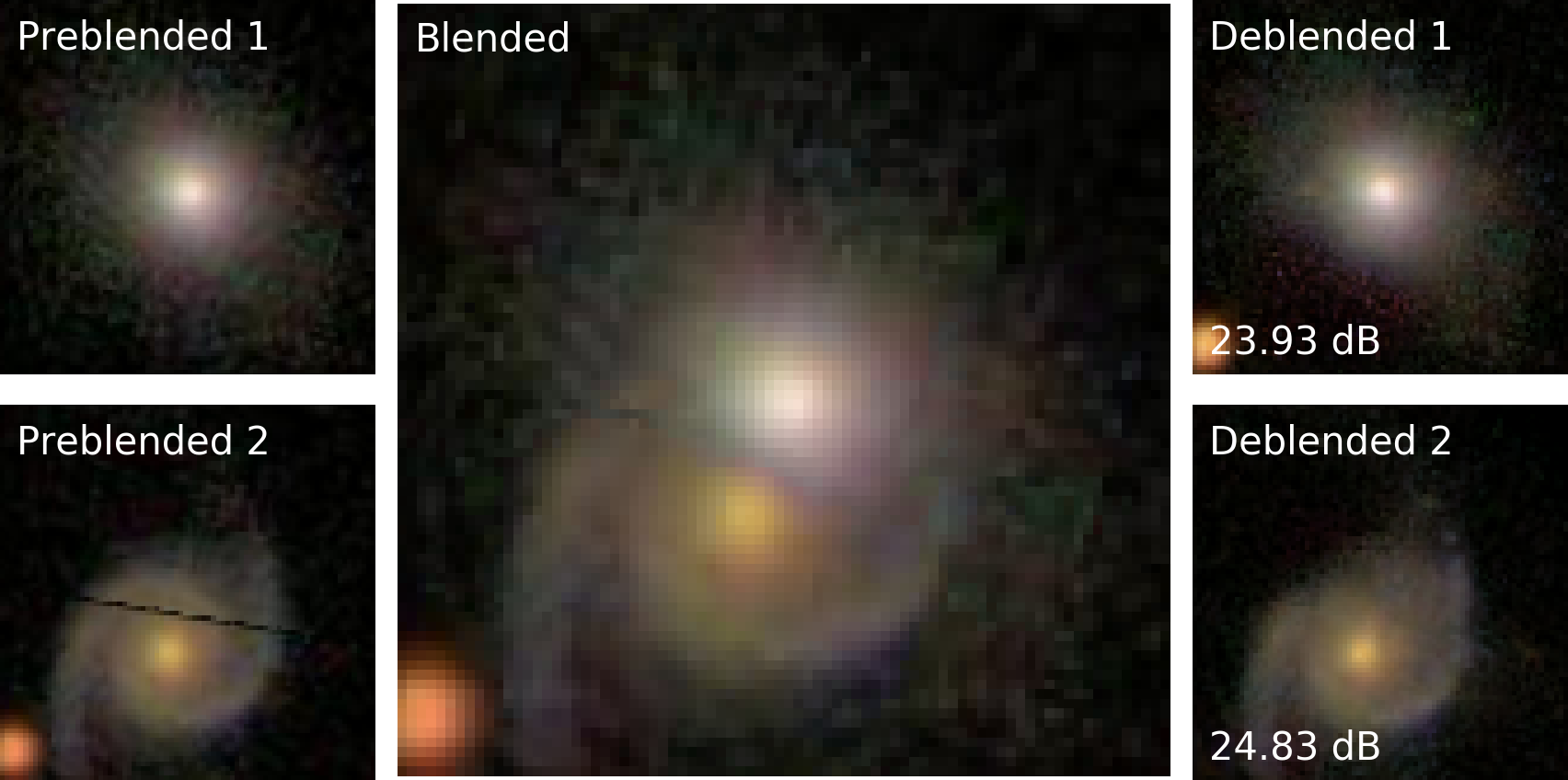}
    \caption{A selection of failed deblender GAN predictions. There exists a variety of failure modes in the deblender GAN predictions: (1) incomplete deblends wherein artifacts of one galaxy are left in the prediction for the other galaxy. (2) Incomplete galaxy generation due to our blending prescription. In selecting the pixelwise maximum value during the blending process, foreground galaxies that eclipse bright regions like the core of the background galaxy often are non-realistically ``cutoff'' by the overwhelming background flux. This leaves our network very little with which to predict from and often results in predictions that appear to be only a part of the associated preblended galaxy. (3) Associating background galaxies with the incorrect blended galaxy. This negatively biases the PSNR scores for otherwise acceptable predictions---a representative example is given in the last image of the above figure. (4) Deblending artifacts: in a handful of predictions, we notice random artifacts of unnatural color such as the green aberration in the second-to-last image above; a possible explanation being that a bright celestial object in the same relative position to the galaxy appears in neighboring galaxy images of the galaxy image manifold.}
    \label{fig:failures}
\end{figure*}
\clearpage

The deblender model presented herein would likely benefit from multiband input data spanning a larger range of the electromagnetic spectrum. Deep neural networks have proven powerful tools for estimating photometric redshifts from multiband galaxy images alone with no manual feature extraction \citep{DeepPhotoZ}. The ability to learn approximate photometric redshift estimates from input images would likely strengthen the deblending performance of our network. To this end, future iterations of our deblender GAN will utilize multiband inputs.

Moreover, our model is restricted to inference on solely nearby (low-redshift) galaxies. In deep learning, we assume that examples from the training and testing sets are drawn from the same distribution. For our deblender, this limits feasible test images to those composed of galaxies from the same redshift and mass bin as the training images. It is commonly understood that galaxy morphology evolves with redshift and therefore galaxy images at high-redshift will look largely different than nearby galaxies \citep{MorphEvolution}, i.e. they belong to a different image manifold. Though the galaxy distributions are distinct, the distribution of pixel intensities in the images share a great deal of similarities such as dark backgrounds populated by luminous, diffuse objects. This observation makes it an area of interest apt for transfer learning which is a topic we will explore in future work.

Also, in future work we will apply our branched deblender GAN to data generated from GalSim \citep{GalSim} in which we can robustly blend any number of galaxies and apply a variety of PSFs. With this, we hope to generalize our results to galaxy images from a wide variety of potential surveys and thereby salvage an increasing number of blended images which would otherwise be unused in the study of galaxies and galaxy evolution. Moreover, we plan to implement a variable number of GAN branches, allowing any blended object to be deblended after estimating the number of objects involved in the blend. An ultimate goal is the full deep learning pipeline from source identification to blending classification and deblending.

\section*{Acknowledgements}

The authors would like to thank Shawfeng Dong for offering his compute resources to train and test our model. We would also like to thank Joel Primack and Doug Hellinger for their helpful feedback on a draft of this paper. In addition, we thank David Koo, Sandra Faber,  Tesla Jeltema, and Spencer Everett for many insightful discussions and recommendations.

The authors would also like to acknowledge the Sloan Digital Sky Survey for providing the data used herein.

Funding for the SDSS and SDSS-II has been provided by the Alfred P. Sloan Foundation, the Participating Institutions, the National Science Foundation, the U.S. Department of Energy, the National Aeronautics and Space Administration, the Japanese Monbukagakusho, the Max Planck Society, and the Higher Education Funding Council for England. The SDSS Web Site is http://www.sdss.org/.

Finally, we'd like to thank the Galaxy Zoo team for their efforts in providing a neatly curated dataset of SDSS galaxy images to the astronomy community.




\bibliographystyle{mnras}
\bibliography{mnras} 








\bsp	
\label{lastpage}
\end{document}